\documentclass[prd,twocolumn,nofootinbib,showpacs,superscriptaddress]{revtex4}

\usepackage{amsmath}
\usepackage{latexsym}
\usepackage{graphicx}

\begin{document}

\title{Binary black hole merger dynamics and waveforms}

\author{John G. Baker}
\affiliation{Gravitational Astrophysics Laboratory, 
NASA Goddard Space Flight Center, 8800 Greenbelt Rd., 
Greenbelt, MD 20771, USA}
\author{Joan Centrella}
\affiliation{Gravitational Astrophysics Laboratory, 
NASA Goddard Space Flight Center, 8800 Greenbelt Rd., 
Greenbelt, MD 20771, USA}
\author{Dae-Il Choi}
\affiliation{Gravitational Astrophysics Laboratory, 
NASA Goddard Space Flight Center, 8800 Greenbelt Rd., 
Greenbelt, MD 20771, USA}
\affiliation{Universities Space Research Association, 
10211 Wincopin Circle, Suite 500, Columbia, MD 21044}
\author{Michael Koppitz}
\affiliation{Gravitational Astrophysics Laboratory, 
NASA Goddard Space Flight Center, 8800 Greenbelt Rd., 
Greenbelt, MD 20771, USA}
\author{James van Meter}
\affiliation{Gravitational Astrophysics Laboratory, 
NASA Goddard Space Flight Center, 8800 Greenbelt Rd., 
Greenbelt, MD 20771, USA}

\date{\today}

\begin{abstract}

We study dynamics and radiation generation in the last few orbits and merger
of a binary black hole system, applying recently developed techniques for simulations of moving black holes.
Our analysis of the gravitational radiation waveforms 
and dynamical black hole trajectories produces a consistent picture for a set of simulations with black 
holes beginning on circular-orbit trajectories
at a variety of initial separations.  We find profound agreement at the level of 1\%
among the simulations for the last orbit, merger and ringdown.
We are confident that this part of our waveform result accurately
represents the predictions from Einstein's General Relativity for the 
final burst of gravitational radiation resulting from the merger of an 
astrophysical system of equal-mass non-spinning black holes. 
The simulations result in a final
black hole with spin parameter $a/m=0.69$.
We also find good agreement at a level of roughly 10\% for the 
radiation generated in the preceding few orbits.

\end{abstract}


\maketitle

\section{Introduction}
\label{sec:introduction}

Two black holes in a binary system spiral together due to the emission
of gravitational waves.  The final merger stage of a binary in
which the black holes have comparable masses
will produce a spectacular burst of gravitational radiation and is
expected to be one of the brightest sources in the gravitational
wave sky.  Mergers of binaries containing two stellar black holes
 are important targets for the first-generation LIGO
gravitational wave detectors, now operating at design sensitivity 
in a year-long science data-taking run, as well as other ground-based
detectors such as VIRGO and GEO. 
Knowledge of the waveforms from the final merger phase is
important to improve the detectability of these sources by 
such detectors\cite{Buonanno00a,Damour:2000gg}.  Mergers of
massive black hole binaries are important sources for 
 the space-based LISA, currently in the formulation
phase. Since LISA is expected to observe these massive black hole
mergers at relatively high signal-to-noise ratios \cite{Flanagan97a}, 
comparison
of the data with calculated merger waveforms should allow a test
of General Relativity in the dynamical, nonlinear regime.

In the early stages of the binary inspiral, the black holes are widely
separated and the waveforms can be calculated analytically using 
perturbative methods.  However, waveforms from the final merger, in which
the black holes plunge together and form a single, highly-distorted black
hole with a common horizon, demand full 3-D numerical
relativity simulations of the full Einstein equations. This proved be a very
difficult undertaking and, for roughly the past decade, 3-D numerical
relativity simulations have been beset by pernicious numerical
instabililties that prevented the simulation codes from running long enough
to evolve any significant fraction of a binary orbit. 

Recently, however, dramatic progress has been made in evolving the merger of
equal mass binary black holes.  Using excision to remove the singular
regions within the horizons and a corotating coordinate system to keep the
black holes fixed in the numerical grid, a binary has been evolved through a
little more than an orbit, and through merger
\cite{Bruegmann:2003aw,Alcubierre:2004hr,Diener:2005mg}, 
though without being able to extract
gravitational waves.  Simulations of excised black holes allowed to move
through the grid on one or more orbits and then through merger have been
carried out, with the extraction of gravitational radiation
\cite{Pretorius:2005gq}.  In addition, new techniques allowing the black
holes to move without the need for excision have been developed
independently by the authors of this paper and another research group
\cite{Baker:2005vv,Campanelli:2005dd}; these have been applied to study the
final plunge, and merger of a binary, with the calculation of gravitational
waves. Recently, these techniques have been applied to evolve a binary with
nonequal masses \cite{Herrmann:2006ks} and the last orbit and merger of an
equal mass binary \cite{Campanelli:2006gf}.  It is especially noteworthy
that this progress is occurring on a broad front, by several independent
groups using different techniques.

Several major open questions in the area of binary black hole mergers center
on the dependence of the resulting gravitational waveforms on the initial
data.  In order to use numerical relativity simulations to compare with data
from gravitational wave detectors, we need to model astrophysically
realistic binary black hole configurations. For non-spinning equal mass
black holes, how strongly do the gravitational waveforms for the merger
depend on the initial data?  What are the effects of spin and non-equal
masses on the resulting waveforms?  The answers to such questions can only
be approached using an evolutionary analysis, in which different initial
data sets are evolved and the resulting waveforms are compared.

In this paper, we take a step towards answering these questions by evolving
several initial data sets for non-spinning equal mass black holes through
the final few orbits, plunge, merger and ringdown. To accomplish this, we apply our
new methods that allow puncture black holes to move freely across a grid.
Using adaptive mesh refinement, we can resolve the dynamical regions near
the black holes (having length scales $\sim M$, where $M$ is the total mass
of the system and we use $c = G = 1$) and the outer regions where the
gravitational waves are extracted (having length scales $\sim (10 - 100)
M$).  Putting the outer boundary typically at $t \approx 768M$, causally
disconnected from the dynamical regions and wavezone throughout most of the
simulation, we can evolve stably for $t > 800 M$. The initial data sets are
chosen to be puncture data~\cite{Brandt97b}. 

Our study focuses on simulations beginning from a set of four inspiralling
black hole configurations modelled as described in Sec.~\ref{sec:init}
and using techniques discussed in Sec.~\ref{sec:techniques}.
In Sec.~\ref{sec:gauging} we calibrate the performance of our
numerical techniques with a resolution study of simulations from the
configuration with the shortest initial separation.
Our main results are presented in
Sec.~\ref{sec:analysis} where we comparatively study the radiation
waveforms and black hole trajectories. 
We find a close relationship between the trajectory information and the
waveforms, providing a consistent picture of the black hole dynamics.
The radiation from the final orbit, merger and ringdown agrees to high
precision among the runs from our range of initial configurations, with the
runs from the farthest initial separation producing a promising waveform
approximately through the last three orbits

\section{Initial Data}
\label{sec:init}

We start by setting up initial data for equal mass binary black holes
represented as ``punctures'' \cite{Brandt97b}.  The metric on the initial
spacelike slice is written in the form $g_{ij} = \psi^4\delta_{ij}$, where
$i,j = 1,2,3$, with conformal factor $\psi = \psi_{\rm BL} + u$. The
static, singular part of the conformal factor takes the form $\psi_{\rm BL}
= 1 + \sum_{n=1}^{2} m_n/2 |\vec{r} - \vec{r}_n|$, where the $n^{\rm th}$
black hole has mass $m_n$ and is located at coordinate position 
$\vec{r}_n$.  The nonsingular function $u$ is calculated by solving the
Hamiltonian constraint equation using {\tt AMRMG} \cite{Brown:2004ma}.

The punctures are initially placed on the $y-$axis in the equatorial ($z =
0$) plane.  We need to specify the individual puncture mass $m$, coordinate
position $Y$, and momentum $P$ so that the black holes are on approximately
circular orbits and have no individual spins (they are irrotational). To
accomplish this, we adapt the quasicircular initial data from the QC
sequence given in Ref.~\cite{Baker:2002qf}. These data are based in turn on
the results of Cook (\cite{Cook94}; see also \cite{Baumgarte00a}), who uses
an effective potential method that minimizes the binary system energy while
holding the orbital angular momentum fixed, to produce an approximately
circular orbit.

The parameters and physical quantities for the simulations we ran are shown
in Table~\ref{table:params}.  Here, $Y$ is the initial coordinate position
of each puncture along the $y-$axis, $P$ is the linear momentum of an
individual puncture, and $m$ is the mass of the puncture. $M_0$ is the initial
total ADM mass of the binary and $J_0$ is the initial total angular momentum.
 $L$ is the initial
proper separation of the black holes, approximated by $L=\int_{H_1}^{H_2}
\sqrt{g_{yy}}dy$, where each limit $H_i$ represents a point on the
Schwarzschild horizon associated with mass $m_i$ at position $\vec{r}_i$.

\begin{table}[h]
 \begin{center}
   \begin{tabular}{c c c c c c c}
   \hline
   \hline
   Run  & $\pm Y$& $\pm P$ & $m$ & $M_0$ & $J_0$ & $L/M_0$   \\
   \hline
   $\;$R1$\;$ & $\;$3.257$\;$  & $\;$0.133$\;$  & $\;$0.483$\;$ & $\;$0.996$\;$ & $\;$0.868$\;$ & $\;$9.9$\;$  \\
   R2 & 3.776  & 0.119  & 0.488 & 1.001 & 0.899 & 11.1 \\
   R3 & 4.251  & 0.109  & 0.49  & 1.002 & 0.928 & 12.1 \\
   R4 & 4.77   & 0.101  & 0.492 & 1.003 & 0.959 & 13.2 \\
   \hline
   \hline
   \end{tabular}
 \end{center}
 \caption{Initial data parameters and physical quantities for the runs
  considered in this paper.}
 \label{table:params}
\end{table}

We note that our runs R1, R2, R3, and R4 correspond closely to the
initial data for models QC-6, QC-7, QC-8, and QC-9, respectively, along the
QC sequence.  As shown in Ref.~\cite{Baker:2002qf}, at these relatively wide
initial separations, the parameters for these initial data sets are close to
those derived using post-Newtonian (PN) techniques; see in particular Figs.
26 - 28 in Ref.~\cite{Baker:2002qf}.

\section{Simulation Techniques}
\label{sec:techniques}

We evolve the initial data with the {\tt Hahndol}
code\cite{Imbiriba:2004tp}, which uses a conformal (BSSN)
formulation of Einstein's evolution equations on a cell-centered
numerical grid.  The basic equations are the same as given in
Ref.~\cite{Imbiriba:2004tp}, with the exception that the evolution
equation for the BSSN variable $\tilde{\Gamma}^i$ has been modified for 
stability as suggested in \cite{Yo02a}. In addition, to reduce high
frequency noise associated with refinement interfaces, we 
add some dissipation of the form given in \cite{CarpetFMR}.  

Mesh refinement and parallelization are implemented in our code with 
the {\tt PARAMESH} package \cite{paramesh,parameshMan}.
We use $4^{\rm th}$-order centered
differencing for the spatial derivatives except for the advection of the
shift, which is performed with $4^{\rm th}$-order upwinded differencing. 
The refinement boundary interfaces are buffered with
$4^{\rm th}$-order-interpolated guard cells which, at worst, may introduce
$2^{\rm nd}$-order errors into second derivatives.
 With our current mesh
refinement implementation, our accuracy is limited by spatial finite
differencing error from refinement interfaces. Since
 we do not gain much by using higher order time
integration, we use $2^{\rm nd}$-order time stepping via a three-step iterative
Crank-Nicholson scheme.   Even though, overall, we expect $2^{\rm nd}$-order 
convergence, we have found considerable advantage in using $4^{\rm th}$-order spatial
differencing over $2^{\rm nd}$-order spatial differencing, as measured by the
accuracy and manifest convergence of the Hamiltonian constraint and other quantities.


Traditionally, puncture black holes have been evolved by keeping them fixed in the
grid \cite{Alcubierre:2004hr,Diener:2005mg}.  This is 
accomplished by factoring out the singular part $\psi_{\rm
BL}$ and handling it analytically, while evolving only the regular parts of
the metric.  As explained in detail in~\cite{Baker:2005vv}, we employ
instead newly developed techniques that allow the puncture black holes to
move freely across the numerical grid.  The singular part is not factored
out; instead the entire conformal factor is evolved. Initially, the binary
is set up so that the centers of the punctures are not located at grid
points (as in the traditional implementation).  Taking numerical derivatives
of $\psi_{\rm BL}$ causes an effective regularization of the puncture
singularity through the inherent smoothing of finite differences. 

Our new approach allows the punctures to move freely by implementing a
modified version of the Gamma-freezing shift vector.  The Gamma-freezing
condition generally improves numerical stability by evolving the coordinates
towards quiescence, in accord with the physical dynamics.  Our modification
to this gauge is tailored for moving punctures, as mentioned in
\cite{Baker:2005vv}. Specifically we use
\begin{equation}
 \partial_t \beta^i = \frac{3}{4} \alpha B^i
\end{equation} 
and
\begin{equation} 
\partial_t B^i = \partial_t \tilde{\Gamma}^i -\beta^j
 \partial_j\tilde\Gamma^i - \eta B^i
\end{equation}
for the shift $\beta^i$ and
\begin{equation}
 \partial_t \alpha = -2\alpha K + \beta^i\partial_i\alpha 
\label{eqn:lapse} 
\end{equation}
for the lapse $\alpha$.  The lapse equation~(\ref{eqn:lapse})
is the well-known ``1+log"
slicing condition, modified with an advection term as in~\cite{Campanelli:2005dd}. 
We use the initial gauge
conditions $\beta=0$, $B=0$, and $\alpha=\psi^{-2}$, similar to those
recommended in~\cite{Campanelli:2005dd}.
We place the punctures in the $z=0$ plane and impose
 equatorial symmetry throughout.

We use adaptive mesh refinement to produce a numerical grid having 
appropriate resolution in the strong-field dynamical regions near the black
holes and in the wave zone. Initially, we set up the black hole binary in a
numerical domain with a box-in-box refinement structure having an innermost
refinement of $h_f$ and subsequent boxes of twice that resolution. We start
with two boxes centered on each individual black hole, and then a box
centered on the origin that encompasses both black holes. Subsequent boxes
centered on the origin are used to give up to a total of 10 refinement
levels.  

During the evolution the black holes move freely across the grid, changing
the curvature in the surrounding region; in response, the initial grid
structure is changed adaptively. Paramesh works on logically Cartesian, or
structured, grids and carries out the mesh refinement on grid blocks.  If
the curvature reaches a certain threshold (a free parameter in our code) at
one point of a block, that block is bisected in each coordinate direction to
produce 8 child blocks, each having half the resolution of the parent block.
If all points in all the child blocks fall below the threshold, those blocks
get derefined.

The box stretching from $-48M$ to $+48M$ in the $x-$ and $y-$directions
and from $0$ to $+48M$ in the $z-$direction is fixed
in place throughout all runs.\footnote{For convenience, we use
$M = 1$ to set the scale for the computational grid and time.  Note that
the actual initial total ADM mass for each case is $M_0 \approx M$,
as given in Table~\ref{table:params}.}  Boxes inside this one can, and generally
do, change adaptively as the simulation evolves; boxes outside this one
maintain their original locations on the grid.  At the initial time,
there is a box stretching out to $24M$; as
the binary evolves, this box generally shrinks as the black holes
spiral in towards the center.  The resolution in the region between
this box and the next (fixed) one at $\pm 48M$ is $h_w = 32 h_f$ in
all runs.  We generally extract gravitational waves on a sphere
of radius $r_{\rm ex} = 30M$.  In the early stages of the run, this
sphere intersects the next innermost box with resolution $h_w/2$;
at later times, it is completely located within the region having
resolution $h_w$.

We extract gravitational waves from our simulations using the Weyl tensor
component $\Psi_4$. Our wave extraction techniques are based on Misner's
method~\cite{Misner:1999ab} and are $2^{\rm nd}-$ order accurate.
They are robust and accurate even when the
extraction radii cross mesh refinement boundaries~\cite{Fiske:2005fx}. In
particular, the waveforms computed at various extraction radii $r_{\rm ex}$
are preserved up to the leading order $1/r$ scaling and show no ill effects
from passing through one or more refinement boundaries.

\section{Calibration of Simulations}
\label{sec:gauging}

We have performed detailed studies of the errors and convergence
behavior of simulations run on the {\tt Hahndol} code using fixed
mesh refinement in previous work \cite{Choi:2003ba,Imbiriba:2004tp,
Fiske:2005fx,Baker:2005vv}.  The simulations presented here differ
from this earlier work in two important ways: they are carried out
using an adaptive mesh structure that changes as the binaries 
evolve, and they are run for significantly longer durations.  In
this section, we discuss the calibration tests we have carried out
to verify that the code produces robust, reliable results in 
these more challenging regimes.

We performed these calibration tests on run R1, which we ran at three
different resolutions.  The initial data parameters for R1 are given in
Table~\ref{table:params} and the simulation parameters for these three cases
are shown in Table~\ref{table:r1}. We use a base resolution of $\rho=3M/32 =
0.09375M$ from which we reach the three resolutions used in the runs. Note
that the medium resolution case,
\begin{table}[h]
 \begin{center}
   \begin{tabular}{c|c c c}
   \hline
   \hline
   R1 cases & low & medium & high   \\
   \hline
   $h_f$ & $\rho/2$  & $\rho/3$ & $\rho/4$  \\
   outer boundary & $\;\pm 768M\;$ & $\;\pm 192M\;$ & $\;\pm 768M\;$ \\
   $T_{\rm sim}$  & $\;186M\;$  & $\;332M\;$ & $\;291M\;$  \\
   \hline
   \hline
   \end{tabular}
 \end{center}
 \caption{Parameters for the low, medium, and high resolution runs of
 model R1, where $\rho = 3M/32 = 0.09375M$. $T_{\rm sim}$ is the
total duration of the simulation.}
 \label{table:r1}
\end{table}
$h_f = \rho/3$, has the outer boundary located relatively close, at 
$\pm 192M$; this was one of the earliest runs we did.
Since the simulation runs for a total duration $T_{\rm sim} = 332M$, small errors
from this outer boundary do have time to propagate in to the physically
interesting regions of the grid.  While these effects are generally
small and have no significant impact on the dynamics or wave extraction,
 we did use a more distant outer boundary at $\pm 768M$ in all
other runs to eliminate this problem. This adds only a small overhead
to the overall cost of the simulation, due to the fixed mesh refinement
structure used in the outer regions.  

\begin{figure}[t]
  \includegraphics*[width=18pc,height=14pc]{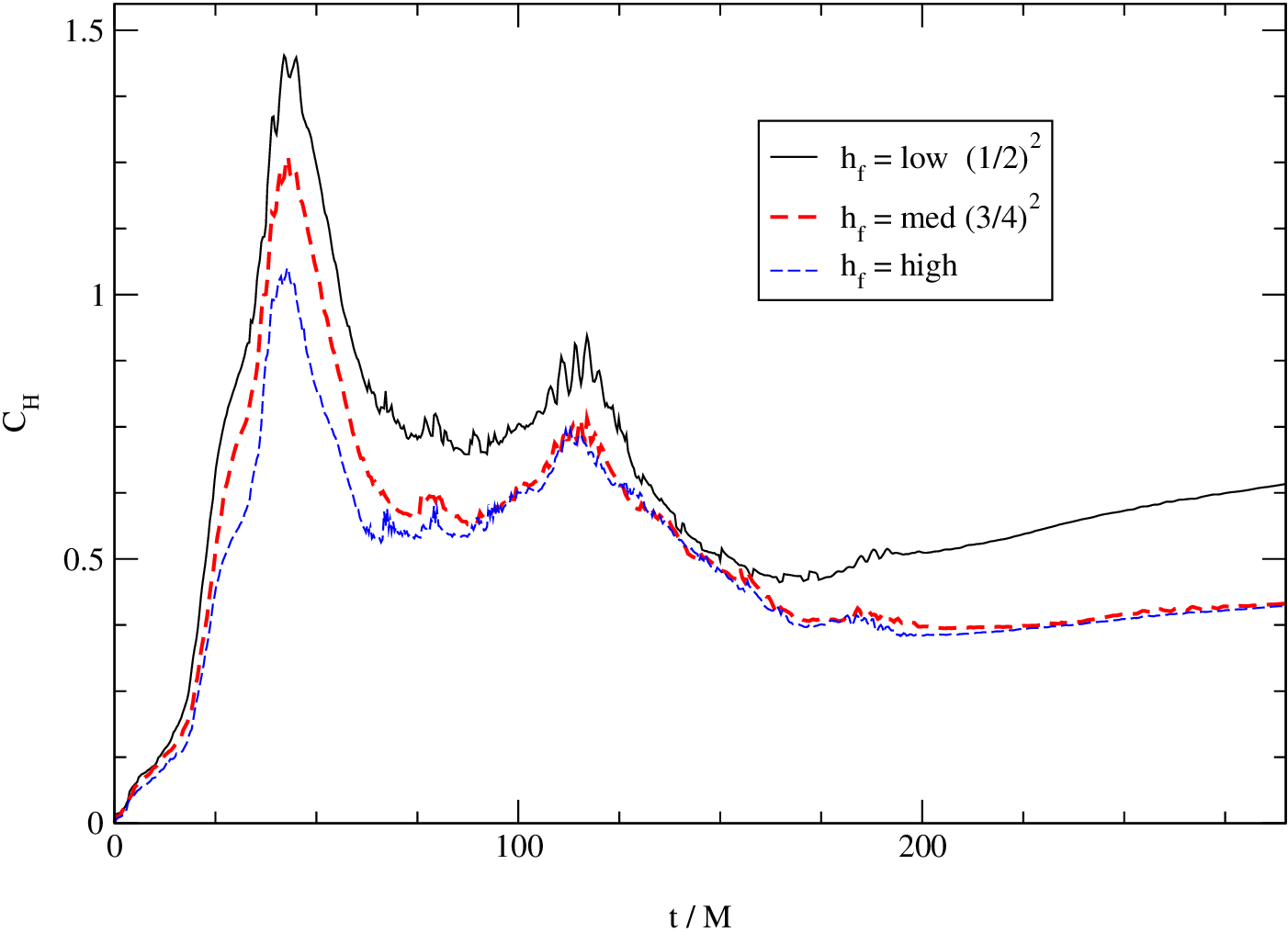} 
  \caption{The L1 norm of the Hamiltonian constraint violation is shown as a
  function of time for the three different resolution runs of $R1$ given in
  Table~\ref{table:r1}.  The high resolution case is shown with a
  dashed line.  The medium (bold dashes) and low (solid) cases are scaled so
  that, for $2^{\rm nd}-$order convergence all three curves would lie on top
  of each other. The L1 norm is taken over all levels of the grid inside
  48M, including the wave extraction region.
  This figure indicates satisfactory
  convergence of the Hamiltonian constraint error in our simulations.} 
  \label{fig:HamConverge}
\end{figure}

For these three runs, we scaled the criterion for refinement and 
derefinement to provide as closely as possible the same grid structure at the
same physical time in the simulations. Since it is generally not possible to match the
grid structures exactly, pointwise convergence tests are of questionable value.
Instead, we calculated the $L1$ norm of the Hamiltonian constraint $C_H$
over the grid as a function of time.  This norm was taken over all levels
inside the box at $\pm 48M$, which includes the wave extraction zone.
Figure~\ref{fig:HamConverge} shows the $L1$ norms for these three runs; note
that the initial growth of the constraint violation is brought to a halt
after approximately $50 M$ of evolution and then diminishes. We attribute
this behavior in $C_H$ to a gauge wave pulse that we have observed leaving
the source region early in the simulation.  The gauge wave has strong high
frequency components and is thus prone to generating differencing error and
reflections from refinement boundaries.  $C_H$ settles down somewhat after
the gauge wave leaves the grid, as suggested by the plot.
The curves are scaled so that, for $2^{\rm
nd}-$order convergence, they would lie on top of each other. As
Fig.~\ref{fig:HamConverge} shows, we get $2^{\rm nd}-$order convergence (or
slightly better) for the entire course of the run. 

\begin{figure}[t]
  \includegraphics*[width=18pc,height=14pc]{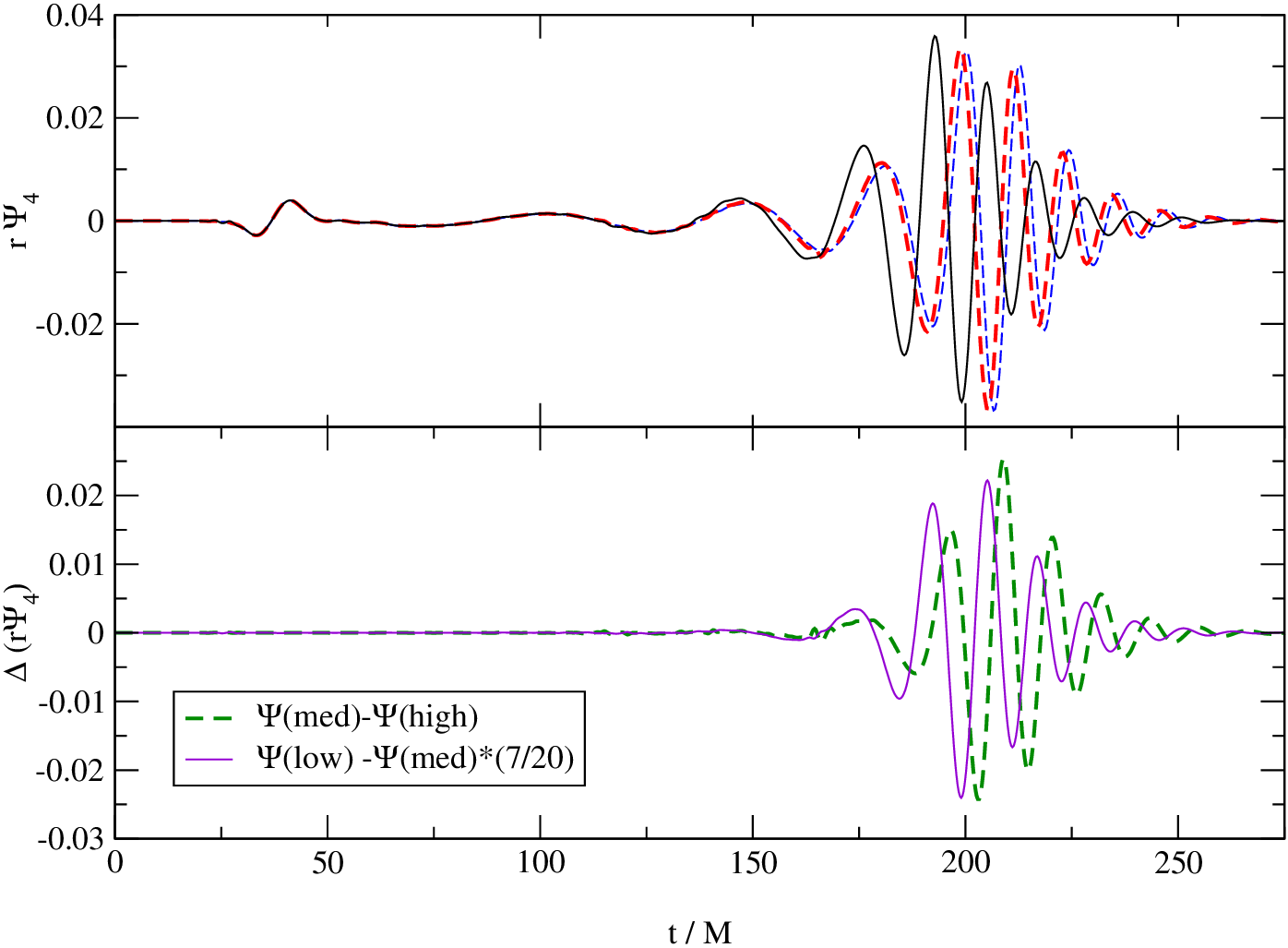}
  \caption{Gravitational waveforms and a naive convergence test.
  The top panel shows
  the $l=2$, $m=2$ mode of $\Psi_4$ for the low (solid), medium (bold dashes),
  and high (dashes) resolution
  runs of R1.  The lower panel shows the differences between these
  waveforms; for $2^{\rm nd}-$ order convergence, the curves would lie on
  top of each other.  Phase differences between the waveforms account for
  the large differences shown.  When the phases are shifted appropriately,
  the convergence of the waves is more manifest, as in Fig.~\ref{fig:WaveConvergeII}}
  \label{fig:WaveConvergeI}
\end{figure}

\begin{figure}[t]
  \includegraphics*[width=18pc,height=14pc]{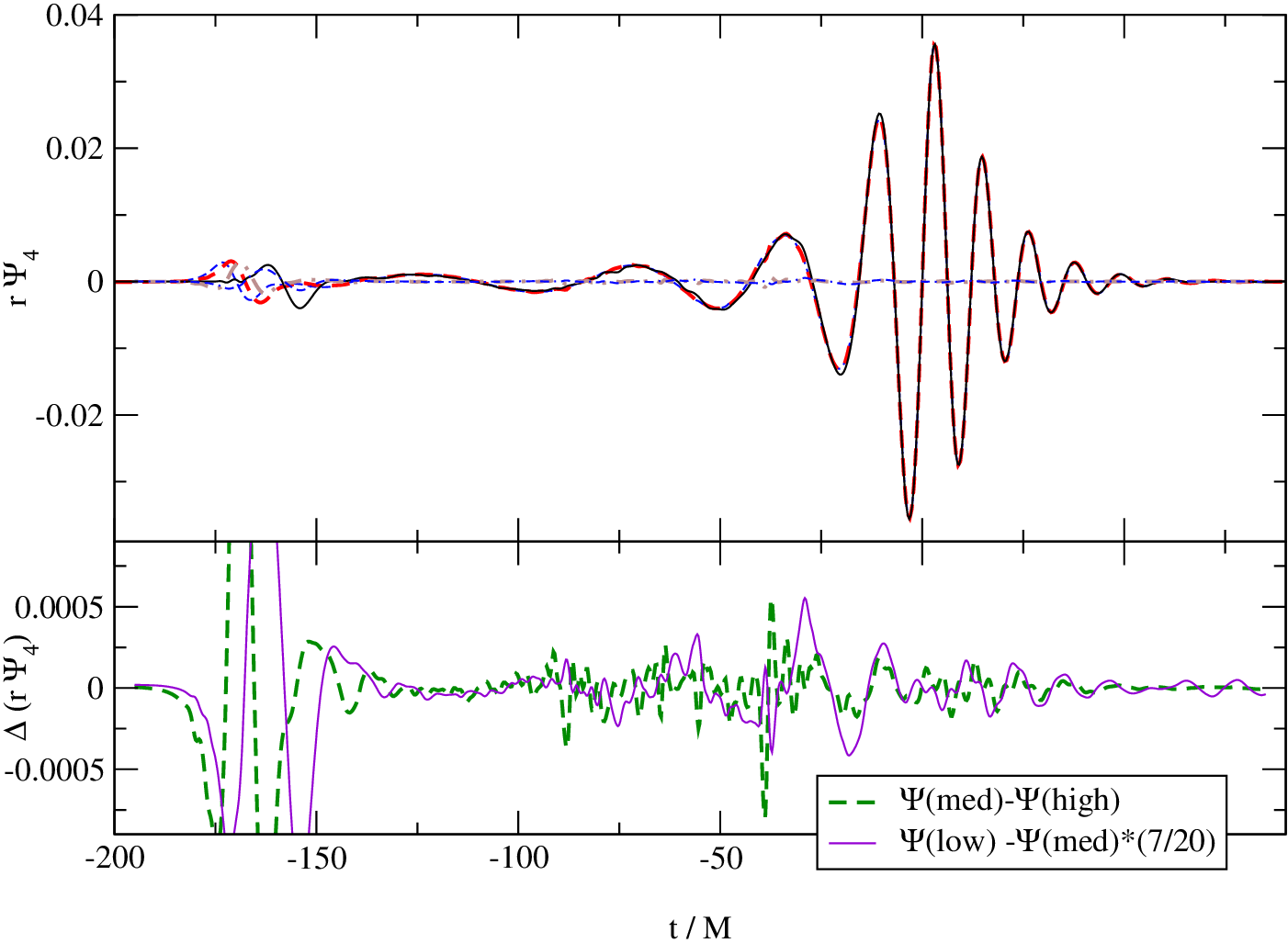} 
  \caption{Time-shifted gravitational waveforms and a physical convergence
  test. The labels in the top panel are as in Fig.~\ref{fig:WaveConvergeI}. In
  the top panel, the gravitational waveforms have been shifted in time so
  that the peak amplitude of the radiation occurs at $t = 0$. The
  differences between these time-shifted waveforms are shown in the bottom
  panel; these curves are scaled so that they would lie on top of each other
  for $2^{\rm nd}-$ order convergence. }
  \label{fig:WaveConvergeII}
\end{figure}

The gravitational waves are extracted on a sphere of radius $r_{\rm ex}
= 30M$.  The top panel of Fig.~\ref{fig:WaveConvergeI} shows the 
$l=2$, $m=2$ component of $r\Psi_4$ for the three different
resolution runs, where $r$ is the estimated areal radius
of the extraction sphere \cite{Fiske:2005fx}.  
The waves start out in phase; note in particular the agreement
in the initial pulse around $t \sim 30M$. However, as the
runs proceed, timing differences slowly accumulate over the
relatively long orbital time scales. By 
$t \sim 150M$, significant differences have accumulated in
the lowest resolution case (solid line) compared to
the other two runs; we attribute this to the larger 
inherent numerical diffusion in this lowest resolution run.

A naive convergence test, in which we take differences directly between
the waveforms at the same simulation times, is shown in the bottom panel of
Fig.~\ref{fig:WaveConvergeI}; these curves are scaled so that, for
$2^{\rm nd}-$order convergence they would coincide.  Comparing the
top and bottom panels, it is clear that the differences between these
curves  can mainly be attributed to the
differences in the phases of the waveforms. 
At several times, the phases are actually off by
$\pi$ so that the differences between the two waves is twice as large as the
individual waves themselves. 

A more meaningful convergence test is shown in Fig.~\ref{fig:WaveConvergeII}.
Here, we have shifted the waveforms in time and phase to adjust for the timing
differences using the techniques described below.  The time axis has
been relabeled so that the peak of the radiation occurs at $t = 0$.
The top panel shows the $l=2$, $m=2$ mode of 
$r \Psi_4$ for the low (solid), medium (bold dashes), and high
(dashes) resolution cases;  note that the physical parts of the
waveforms ($t \gtrsim -125M$) agree beautifully.  We then carried out a
convergence test using these shifted curves; the results are shown in
the bottom panel.  Again, the curves are scaled so to lie on top of
each other for $2^{\rm nd}-$ order convergence.  Here we see nearly
perfect convergence for the physical parts of the waveforms; note
in particular the much smaller vertical scale in the lower panel
of Fig.~\ref{fig:WaveConvergeII}.

\begin{table}[h]
  \begin{center}
  \begin{tabular}{c| c c c c c}
  \hline
  \hline
   $h_f$ & &$r_{\rm ex}=20M\;$ & $r_{\rm ex}=30M\;$ & $r_{\rm ex}=40M\;$ & $r_{\rm ex}=50M$\\
  \hline
  $\rho/2\;$ &
  $\;E_{rad}\;$   & $0.0347$ & $0.0343$ & $0.0336$ & $0.0324$ \\
   & $J_{rad}$    & $0.217$  & $0.218$  & $0.218$  & $0.215$  \\
  \hline
  $\rho/3$ &
  $E_{rad}$    &  $0.0343$ & $0.0345$ & $0.0345$ &  $0.0343$\\
   & $J_{rad}$   &$0.215$  & $0.223$ &$0.226$ & $0.227$  \\
  \hline
  $\rho/4$ &
  $E_{rad}$    & $0.0342$ & $0.0344$ & $0.0345$ & $0.0344$  \\
   & $J_{rad}$    & $0.216$ & $0.224$ & $0.227$ & $0.228$  \\
  \hline
  \end{tabular}
  \end{center}
  \caption{Values of the energy $E_{rad}$ (in units of $M_0$)
 and angular momentum $J_{rad}$ (in units of $M_0^2$)
   carried away by gravitational radiation 
  for the R1 runs calculated for
  different extraction radii and different resolutions.}
  \label{table:qc6numbers}
\end{table}

As we discussed in~\cite{Baker:2005vv}, the gravitational waveforms can 
be used to calculate the total energy $E_{rad}$ and total angular momentum 
$J_{rad}$ carried away by the radiation. We calculate $dE/dt$ and $dJ/dt$ 
from time integrals of all $l=2$ and $l=3$ waveform components using 
Eqs.~(5.1) and (5.2) in \cite{Baker:2002qf}. 
Integrating $dE/dt$ gives the total energy loss due to the radiation, $E_{rad}$.
We find that the influence of higher modes of the waves contribute 
$< 1\% $ to the energy and even less to the angular momentum.

The total radiated energies calculated from the waveforms extracted at
different radii and with different resolutions are reported in
Table~\ref{table:qc6numbers}. For lower resolution, the radiated energy
depends somewhat on the extraction radius, decreasing with increasing
radius. For the medium and high resolution cases, however, the values are
almost independent of the extraction radius. This indicates that the further out, 
lesser refined regions of the grid produce significant dissipation of
the waves in the low resolution case, whereas when the resolution is high
enough, the lesser refined regions do not have such an effect. 
These considerations indicate that the best radius to extract the energy is 
at $r_{\rm ex}=30M$, in a region refined enough 
that the energy does not significantly dissipate in the low resolution run,
yet far enough way
from the source that it changes only
minimally for higher resolution runs.

As shown in Table~\ref{table:qc6numbers}, the radiated angular momentum
$J_{rad}$ varies by a few percent between $r_{\rm ex} = 30M$ and
$r_{\rm ex} = 50M$ at high resolution.
This observation seems to agree with the
notion that the angular momentum depends more strongly on the longer
wavelength parts of the waves~\cite{Baker:2002qf}, which should be extracted
at greater distances.
For all the
runs reported in this paper, we use an extraction
 radius of $r_{\rm ex} = 50M$ to calculate $J_{rad}$.

\section{Results}
\label{sec:analysis}

In this section we comparatively analyze the results of simulations
based on our R1-R4 black hole configurations.  Recall that the initial
data for each of these simulations starts the black holes at different
separations, with proper separations varying from $9.9M_0$ to $13.2M_0$,
and provided with sufficient angular momentum that the runs are
estimated to be on initially circular orbits.

We can think of each of these simulations as an approximate
representation of the late-time portion of an ideal spacetime which
begins with arbitrarily well-separated black holes on an inspiraling
trajectory which asymptotically approaches circular orbits.  As such
representations, there are several limitations which the initial data
models may have.  In particular,
the initial parameters will only approximate the ideal trajectory since the 
angular momentum as a function of radius may differ from the value 
required for an idealized circular orbit.  Likewise, at finite radius, 
the ideal spiral trajectory can only be approximated by our initially 
circular configurations.  

Furthermore the manner by which we have mapped 
from these trajectory specifications to actual initial data values 
necessarily requires  making some suppositions.  For instance, 
our puncture data prescription, like almost all field prescriptions, 
 contains no representation of prior
radiation generated before the time at which the initial data are posed.
Though it is generally expected that the significance of such limitations on 
the final merger simulations should be reduced if the black holes begin 
sufficiently far apart, there is no clear way to assess
just how significant 
such effects will be on the results, including the gravitational
waveforms, before carrying out the evolutions.

Our simulations, begun with varying initial separations, should be affected 
by any initial modeling error in varying amounts, but should 
agree to the degree that they represent the ideal astrophysical spacetime.
A key objective in our analysis is to identify {\em universal} 
characteristics among the different runs which, we reason, are then 
likely to correctly represent those aspects of the astrophysical 
equal-mass non-spinning binary black hole merger spacetime. 

\subsection{Overview of Simulations}
\label{sec:analysis-overview}
\begin{figure}[t]
  \includegraphics*[width=22pc,height=21pc]{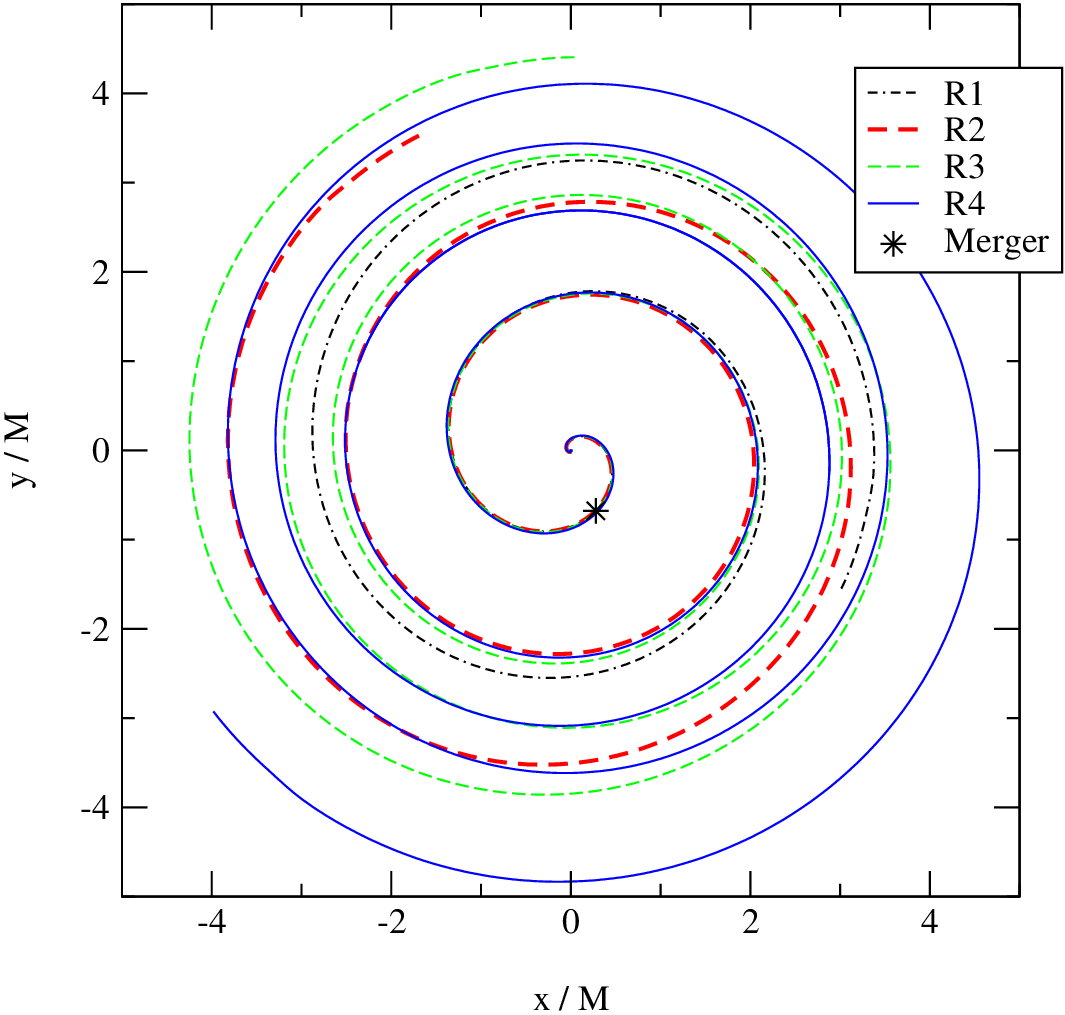}
  \caption{Paths of black holes starting from different initial separations.
   For clarity, we show only the track of one of the black holes from
   each simulation.
  The paths are very similar for approximately the last orbit, indicating
  that the black holes follow the same tracks. The point of merger
  (estimated by a single connected isosurface of $\alpha=0.3$) is indicated
  by as an asterisk in the plot.} 
\label{fig:tracks}
\end{figure}

Our comparative analysis covers four simulations labeled R1 to R4 in
Table~\ref{table:params}. We evolved them all using the medium resolution of
$h_f =\rho/3$ except for R1, where we have applied the higher $h_f =\rho/4$
resolution. In all runs we used an initial grid setup and adaptive mesh
refinement as described in Sec.~\ref{sec:techniques}. We evolved all the
runs to well after the wave signal had passed the extraction region; the
actual amount of time is noted as $T_{\rm sim}$ in
Table~\ref{table:mergertimes}.  For the time-slicing condition used in our
simulations, the region where the lapse satisfies the condition $\alpha=0.3$
corresponds roughly with the apparent horizon location.  We thus used the
moment when the two $\alpha=0.3$ regions around the black holes merge to
specify a merger time $T_{\rm merge}$. The number of orbits for each run,
$N_{\rm orbits}$, was estimated from the trajectories shown in
Fig.~\ref{fig:tracks} and is taken up to the point at which the merger
occurs.
\begin{table}[h]
  \begin{center}
  \begin{tabular}{ c| c c c c}
  \hline
  \hline
  & R1 & R2 & R3 & R4\\
  \hline
  $L/M_0$            & 9.9   & 11.1  & 12.1  & 13.2 \\
  $h_f$  & $\rho/4$  &  $\rho/3$  &  $\rho/3$ &  $\rho/3$    \\ 
  $T_{\rm sim}$     & $421M$    & $531M$    & $530M$    & $850M$   \\
  $T_{\rm merger}$       & $160M$ & $234M$ & $396M$ & $513M$\\
  $N_{\rm orbits}$      & 1.8    & 2.5    & 3.6    & 4.2   \\
  \hline 
  \end{tabular}
  \end{center}
  \caption{Simulation parameters and general results.  $T_{\rm merger}$ is 
the time at which the merger occurs, starting from the initial time
in each run.}
\label{table:mergertimes}
\end{table}

A graphical overview of our four simulations is presented in
Fig.~\ref{fig:tracks} showing the paths traced by the black hole punctures
on the computational domain.  These were obtained by numerically integrating
the equation of motion $\dot{\vec{x}}_{punc}=-\vec{\beta}(\vec{x}_{punc})$, 
which analytically gives the exact trajectory of each
puncture~\cite{Campanelli:2005dd}. The value of the shift at the 
location of the puncture $\vec{\beta}(\vec{x}_{punc})$ was
interpolated between grid points as required.  

For clarity, Fig.~\ref{fig:tracks} shows only the track of 
one of the two black holes from each simulation. We have oriented
each trajectory according to a physical reference discussed
in Sec.~\ref{sec:analysis-raditation}, so that they superpose at the 
radiation peak, which occurs very  near the end of the puncture trajectory.
R4 has the widest initial separation and completes the largest
number of orbits.  Each of the other cases, after an initial
transient period of approximately one orbit, nearly locks on to the
R4 trajectory.  For the final orbit, all four trajectories are very
nearly superposed.
In Sec.~\ref{sec:analysis-trajectories}
we study the quantiative features of these trajectories in more detail.

\begin{figure}[t]
  \includegraphics*[width=18pc,height=14pc]{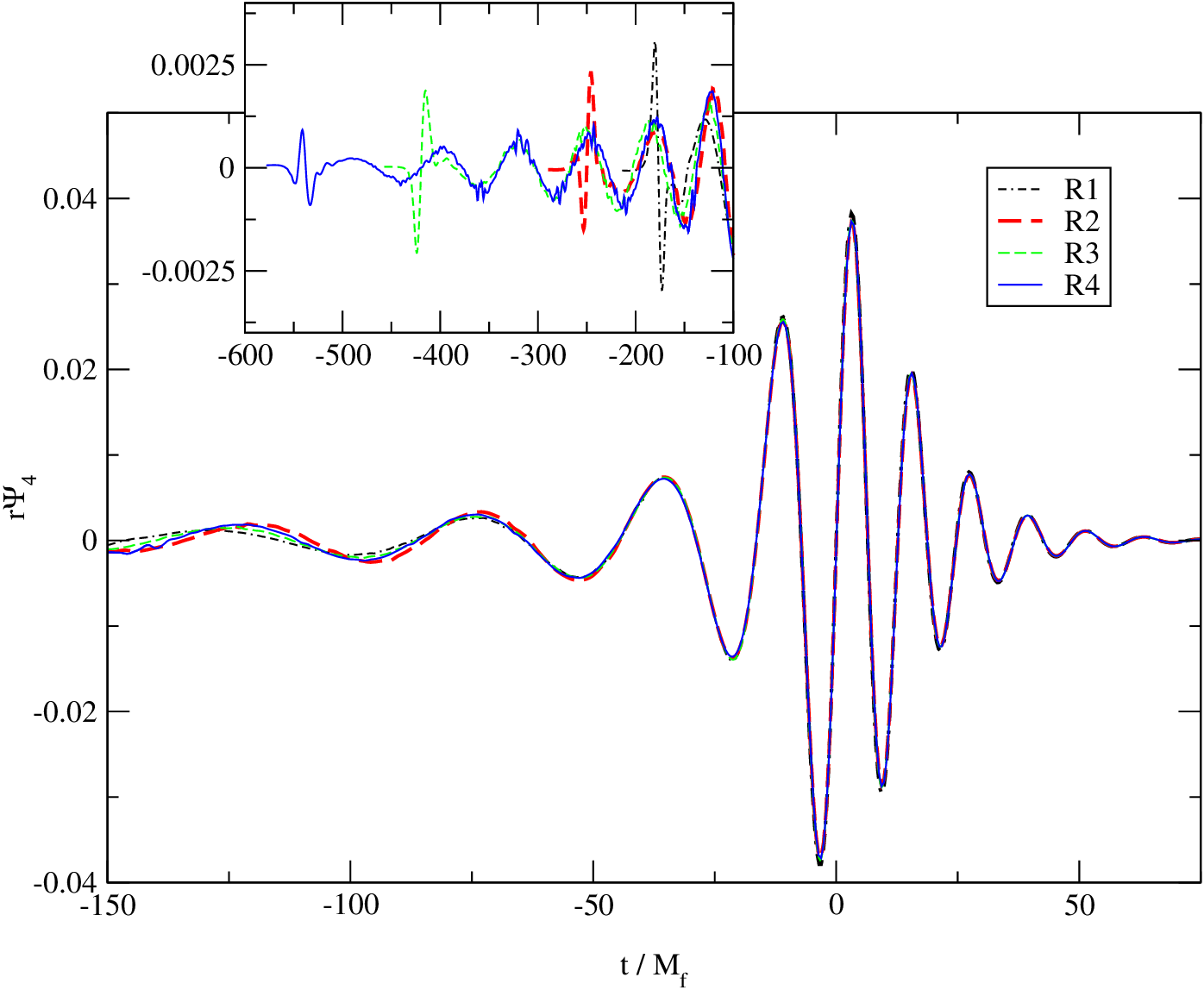}  
  \caption{Waveforms from runs R1 - R4. The figure shows nearly perfect agreement
after $t=-50 M_f$. For the preceding $500M_f$, shown in an inset, the waveforms
agree in phase and amplitude within about 10\% except for a brief initial pulse 
at the beginning of each run. }

  \label{fig:WaveForms}
\end{figure}

Fig.~\ref{fig:WaveForms}
shows one polarization component 
of $r \Psi_4$ for the runs R1 - R4, extracted at $r_{\rm ex} = 30M$
and shifted in time and phase as described below.
Notice that beyond about $t=-50M_f$ the
waveforms superpose sufficiently well that it is not possible to distinguish 
the curves in the plot.  The very strong agreement among our simulations on 
this part of the wave gives us confidence that our waveform accurately 
represent the astrophysical merger-ringdown signal to high precision.
The inset shows that the agreement remains generally good going back to the 
beginning of each simulation, with the R3 and R4 runs agreeing fairly well 
over some $450 M_f$.

\subsection{Gravitational Radiation}
\label{sec:analysis-raditation}

The first step in quantitatively comparing our runs is to calculate the energy and 
angular momentum carried away by the gravitational radiation generated
by the binary system in our simulations.  Following the discussion in 
Sec.~\ref{sec:gauging}, we measure the radiation energy extracted at 
$r_{\rm ex}=30M$, estimating that these will be accurate within $\sim 1\%$.  
We subtract the radiation energy $E_{rad}$ from the initial mass
$M_0$, given in Table~\ref{table:params}, to determine the 
final black hole mass in each simulation, $M_f=M_0-E_{rad}$.  This provides
a physical scaling which we use to compare the R1-R4 simulations.
Similarly, we measure the angular momentum content of the radiation 
$J_{rad}$ as extracted at $r_{\rm ex}=50M$, estimating the accuracy
to be within a few percent.
Subtracting this from the initial ADM angular momentum $J_0$, we
calculate the spin parameter of the final black hole 
$a/M_f=(J_0-J_{rad})/M_f^2$.  
The results are summarized in Table~\ref{table:RN_EandJ}.

\begin{table}[h]
  \begin{center}
  \begin{tabular}{c| c c c| c c}
  \hline
  \hline
   & $E_{rad}/M_f$ & $J_{rad}/M_f^2$ & $a/M_f$ & $M_{QN}/M_f$ & $a_{QN}/M_f$\\
  \hline
  $R1$ &  0.0356 & 0.246 & 0.694 & 1.005 & 0.721 \\  
  $R2$ &  0.0369 & 0.272 & 0.691 & 1.002 & 0.686 \\
  $R3$ &  0.0381 & 0.306 & 0.689 & 1.004 & 0.694 \\
  $R4$ &  0.0387 & 0.325 & 0.702 & 1.004 & 0.693 \\
  \hline
  \end{tabular}
  \end{center}
  \caption{Energy and angular momenta for the radiation and final black hole.
   $E_{rad}$ and $J_{rad}$ are measured at $r_{\rm ex}=30M$, and $r_{\rm ex}=50M$,
 respectively.
   $M_{QN}$ and $a_{QN}$ are calculated independently
from the quasi-normal fits of the ringdown waveforms, and agree well with the
values deduced from the radiative losses.}
  \label{table:RN_EandJ}

\end{table}

We note that the energy and angular momentum content of the radiation is
almost entirely contained in the $l=2$, $m=\pm2$ spin $-2$-weighted
spherical harmonic components, with other components entering at the
1\% level.  In the remainder of our waveform analysis we concentrate
exclusively on the leading component,
\begin{equation}
r\psi_4(\theta,\varphi)\sim 
r\Psi_{4\,(22)}\left({}_{-2}Y_{2\,2}(\theta,\phi)+{}_{-2}Y_{2\,-2}(\theta,\phi)\right)
\end{equation}
where, for simplicity, we have suppressed retarded time dependence.
Hereafter we suppress the multipole labels and refer to the leading
component simply as $r \Psi_4$.  The two polarization components of
the radiation are represented in the real and imaginary parts of 
$r \Psi_4$.  We follow Refs.~\cite{Baker:2002qf,Baker:2003ds}, 
representing our waveforms
by $r \Psi_4=A\exp(-i \varphi_{pol})$, where the amplitude $A$ and
polarization phase $\varphi_{pol}$, like $r \Psi_4$, are functions of
time.  Of course, any complex time-series could be represented in this
form, but it is particularly valuable if the radiation exhibits a {\em
circular polarization} pattern, so that $A$ and $\varphi_{pol}$
vary slowly compared to the wave frequency timescale.  Such radiation
will be circularly polarized to an observer on the system's rotational
axis, varying to linearly polarization for an observer on the
equatorial plane.  We find that the radiation produced in our simulations
shows strong circular polarization, which we will utilize in comparing the 
radiation from our four simulations.

As we are interested in the degree to which our various runs may be taken to
be different models of the same physical merger spacetime, we must define a 
physical basis for comparing them.
As expected, runs beginning with more separated black holes take
a longer time to reach the point of merger.  For each run, a physical reference
time
is recognizable by the point at which the radiation 
reaches its peak amplitude; we define
$t=0$ at this point for our comparisons.

\begin{figure}[t]
  \includegraphics*[width=18pc,height=14pc]{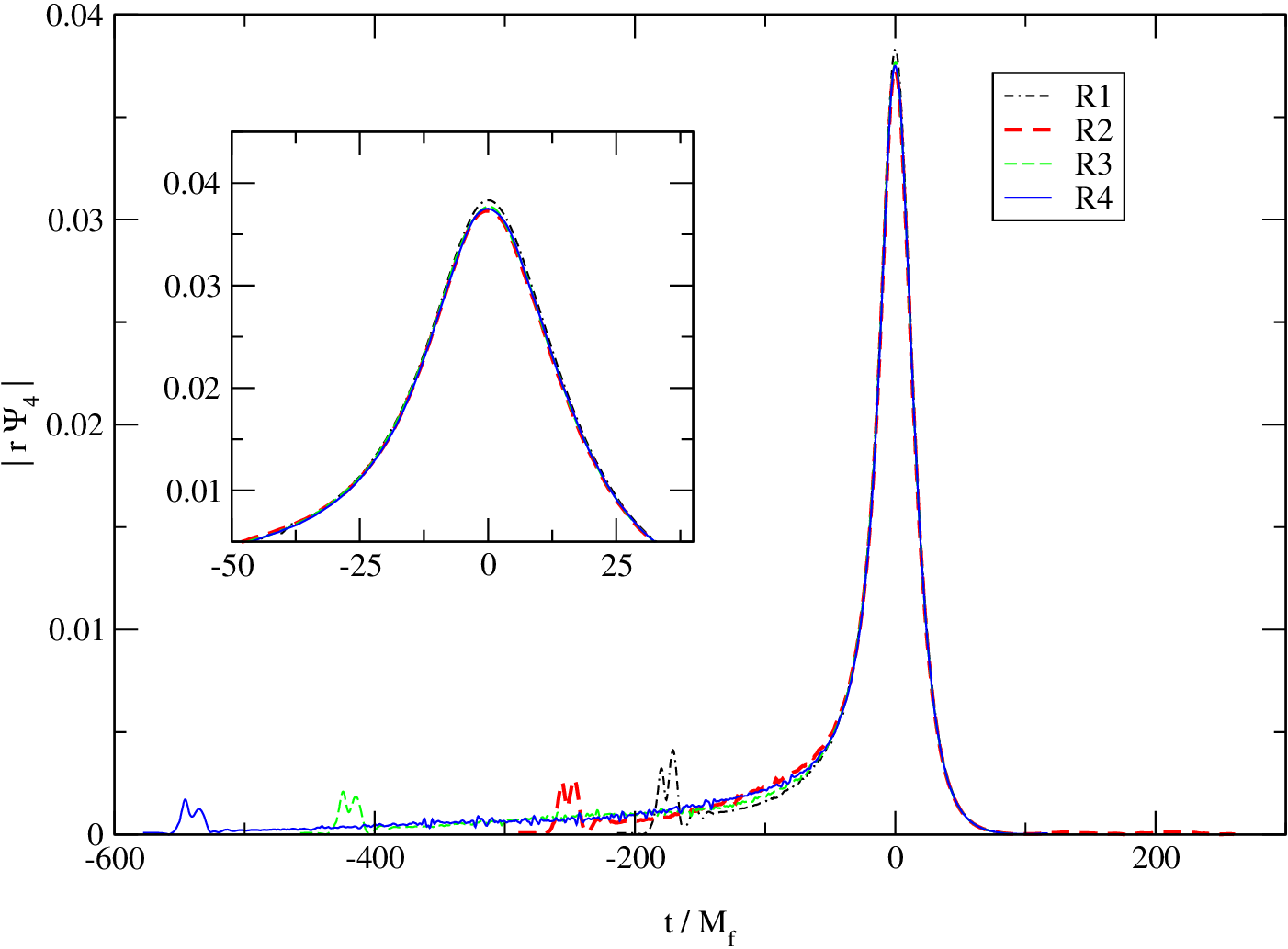}
  \caption{Amplitudes, absolute value of (complex) $\Psi_4$, of the waves.
  The curves have been shifted such that the maxima are all at time 0. The
  inset zooms into the peak showing the strong agreement from $t=-50M_f$ on.
  We have used the amplitude peak as a reference to align
  our simulations in time.}
  \label{fig:WaveAmp}
\end{figure}

Fig.~\ref{fig:WaveAmp} shows the wave amplitudes $A(t)$ from all four runs.
Through the strong radiation peak after $t=-50M_f$ all four wave amplitudes 
show striking
universality in the compared results, with agreement among all runs
to about 1\%.  This period of strong agreement covers roughly the last orbit, plunge and ringdown of the merger.  
Agreement within about 10\% is maintained among the 
R2-R4 simulations for most of each run with slightly more difference in the
R1 run.  The smooth shape of the peak, lacking any sign of the several wave 
cycles spanning the peak (see Fig.~\ref{fig:WaveForms}) is an indication
of the circular polarization pattern discussed above.  The only clearly
non-universal feature in the wave amplitudes is a small and brief 
burst lasting about $50M_f$ at the beginning of each run.  We interpret this 
burst as ``spurious radiation'' content in the initial data.  As generally
expected the amplitude of the spurious radiation lessens as the initial 
separation of the black holes is increased.  

Since the black holes 
in our various runs (which all begin on the $y$-axis) undertake differing 
amounts of orbital motion before merger, we must expect differing orientations 
for the systems at the point of merger.  For meaningful comparsions we must 
rotate the data from each system to align them with repect to some physical 
reference.  As our reference, we will orient the systems so that the 
polarization phase $\varphi_{pol}$ of the radiation passes though zero 
at the moment of peak amplitude $t=0$.  We use this orientation for all
figures with wave phase information, and in the trajectory comparison in 
Fig.~\ref{fig:tracks}. 

\begin{figure}[t]
  \includegraphics*[width=18pc,height=14pc]{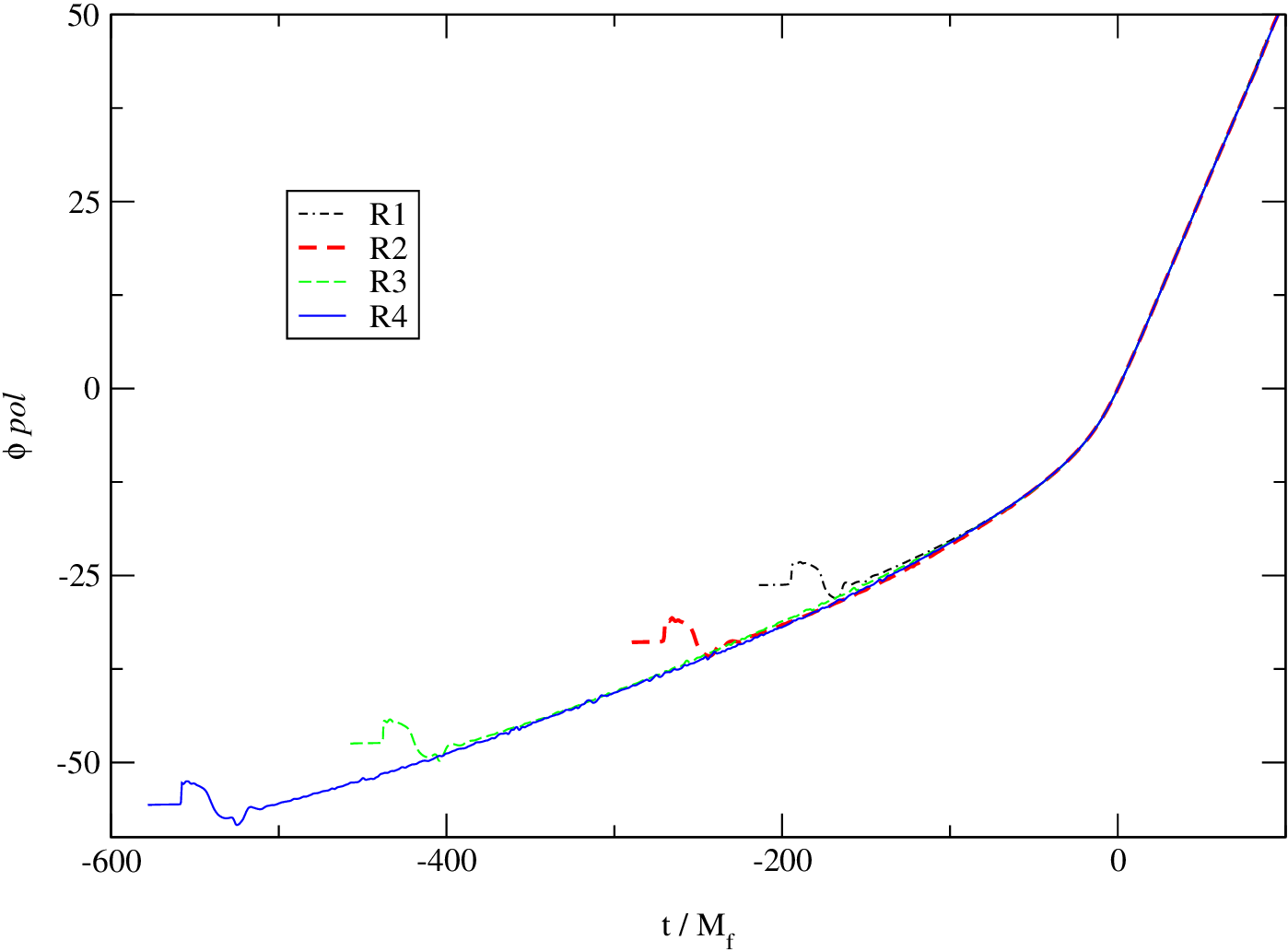}  
  \caption{Graviational wave phase angle vs. time. The phase is made to 
  agree at time $t=0M_f$. After $t=-50M_f$ the phases agree very well 
  for all the runs. For the duration of the runs (except for a brief 
  initial period), the waveforms from all runs agree in phase within about 10\% 
  of a wave cycle.}
  \label{fig:PhaseVTime}
\end{figure}

\begin{figure}[t]
  \includegraphics*[width=18pc,height=14pc]{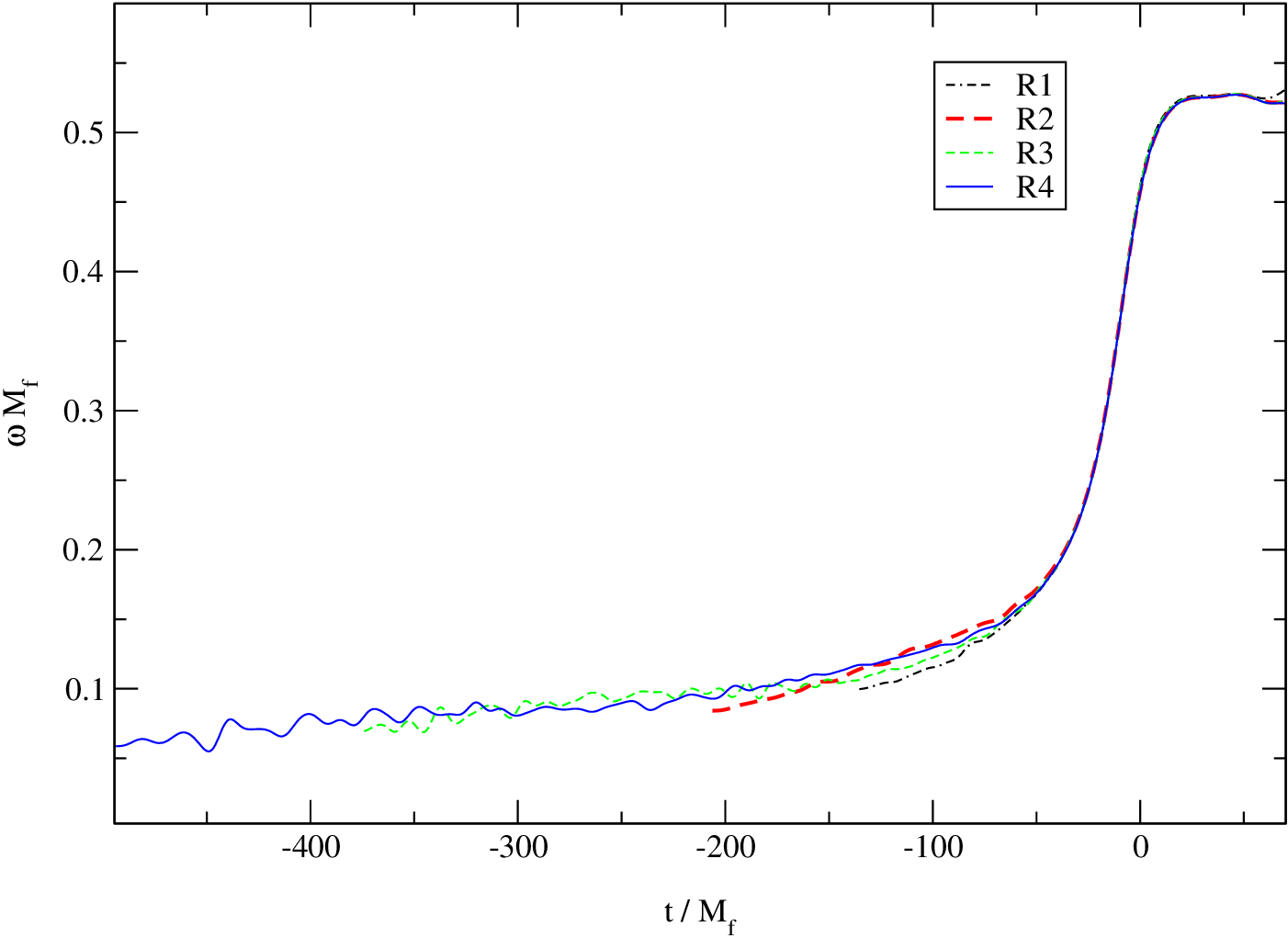}  
  \caption{Waveform frequency as a function of time.  The curves are nearly indisinguishable
 after  $t=-50M_f$.} 
  \label{fig:FreqVTime}
\end{figure}

The polarization phase $\varphi_{pol}$ is shown in Fig.~\ref{fig:PhaseVTime}. 
At time
$t=0M_f$ the phases are set to agree. However, they keep agreeing after that,
showing that the ringdown frequency is the same. 
The plot shows generally good phase agreement among all runs, to within
a small fraction of a wave cycle, except for a brief period at the
beginning of each run when the radiation is dominated by spurious
radiation associated with initial data modeling error, and is not
circularly polarized.  For circularly polarized radiation it is
meaningful to define an instantaneous frequency for the wave,
$\omega=\partial \varphi_{pol}/\partial t$, shown in
Fig.~\ref{fig:FreqVTime},
which allows a more
detailed comparison of the simulation waveforms. It can be seen
that apart from the noise due to the smallness of the waves
initially and at the end, all runs have the same frequency evolution
from about $60M_f$ before the merger.  Before that the frequency evolution
compares similarly among the simulations with some wavyness which may 
correspond to some ellipticity in the early part of each simulation's insprial.

The final (constant) frequency is the frequency of the ringdown. Fitting an
exponential decay to the amplitudes and using this final frequency we
estimate, using the procedure in Ref.~\cite{Leaver86}, the mass and 
angular momentum of the black hole formed in the merger. These estimates 
provide a description of the final black hole as determined by its 
perturbative dynamics.
The results are listed in Table~\ref{table:RN_EandJ}.  where they can 
be compared with the independent estimates obtained by subtracting the 
radiation losses from the initial values.  The good agreement, to $1-2\%$,
provides a measure of energy and angular momentum 
conservation in each simulation.  We also note the strong agreement among
our simulations (which start
 from different separations) on the final state of the remnant
black hole, with the measures from the R2-R4 runs consistent with the 
same value, $a/M_f=0.69(\pm 1\%)$. 
The R1 value differs by a few percent, which may be a consequence of the 
shorter duration of this simulation, allowing greater sensitivity to initial 
transient effects.

 

\subsection{Trajectory analysis}
\label{sec:analysis-trajectories}

In this section we consider the particle-like dynamics of our simulations 
defined by the coordinate trajectories of our black hole punctures.
It is important to be careful in interpreting such coordinate-dependent 
information which may include non-physical gauge features. 
Nonetheless it is worth noting that the Gamma-freezing gauge condition 
applied in our simulations, to the extent that it approximates 
$\tilde\Gamma^i=\tilde\gamma^{ij}_{,j}=0$,
 is similar to the Dirac gauge applied in some 
post-Newtonian calculations, and might be expected to provide a sensible 
coordinate system at least in weak field regions.  
In any case, since we apply similar coordinate conditions in each run, the 
coordinate-based puncture trajectories provide another opportunity to 
identify universal features in our simulations 
that start from different initial 
separations.  

\begin{figure}[t]
  \includegraphics*[width=18pc,height=14pc]{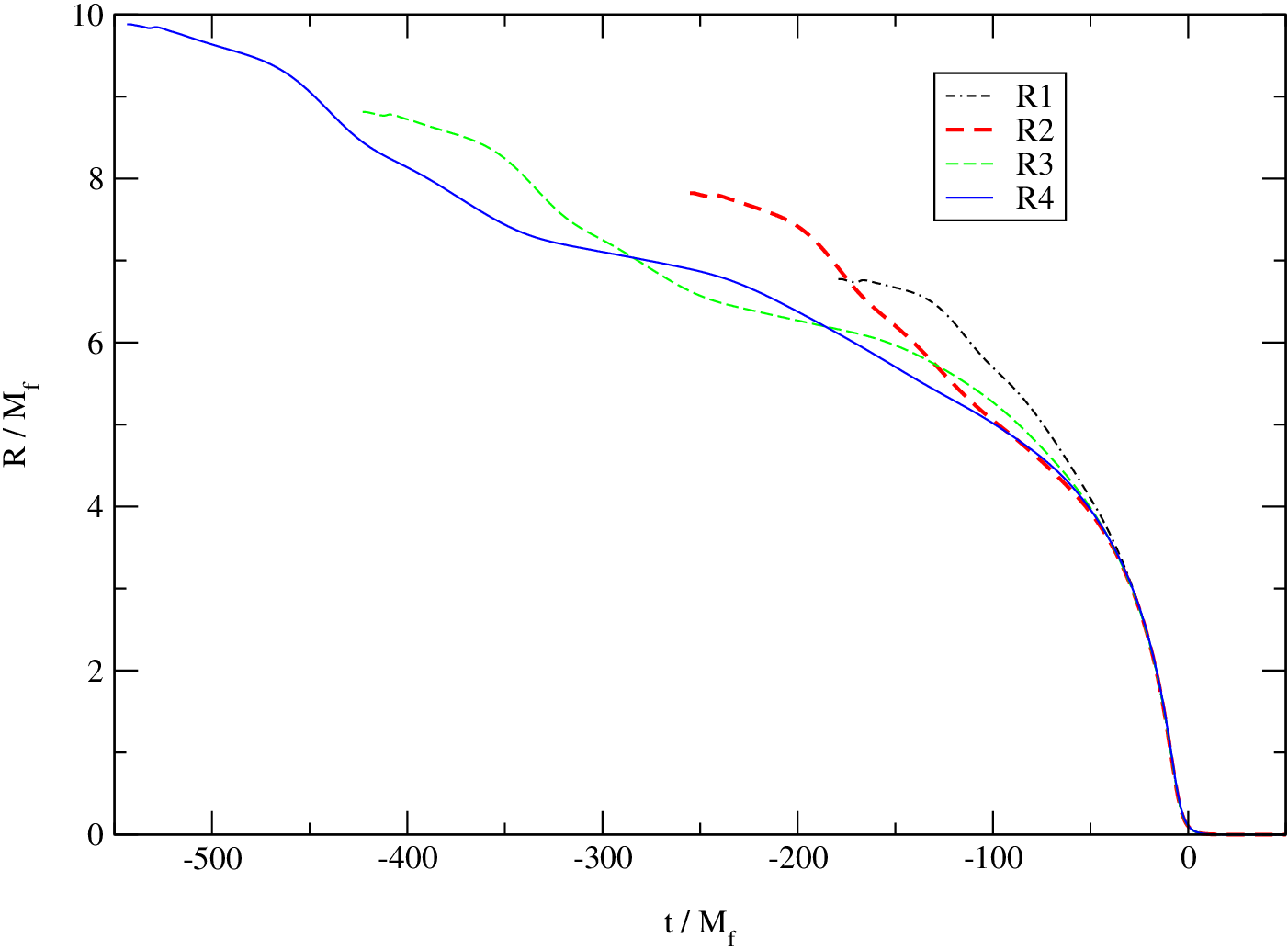}
  \caption{The coordinate separation between the punctures is
shown for runs R1 - R4 as a function of time. Early on in each simulation 
the separation
seems to drop quickly, but all runs track together as they approach merger. }
  \label{fig:rtracks}
\end{figure}

Refer again to Fig.~\ref{fig:tracks}, which shows the tracks of one
of the punctures from each of the runs R1 - R4 as they spiral into
the center.  As the black holes descend deeper into the strong
field, we see that their paths lock on to a universal trajectory
that takes them into the plunge and subsequent merger.
This behavior can also be seen in Fig.~\ref{fig:rtracks},
which shows the coordinate separation between the punctures
as a function of time.
In this case, there is strong agreement among the runs after $t \approx -50M_f$. 
In the earlier part of the simulations there are clear differences among the 
runs, which are suggestive of ellipticity in the initial orbital motion.

\begin{figure}[t]
  \includegraphics*[width=18pc,height=14pc]{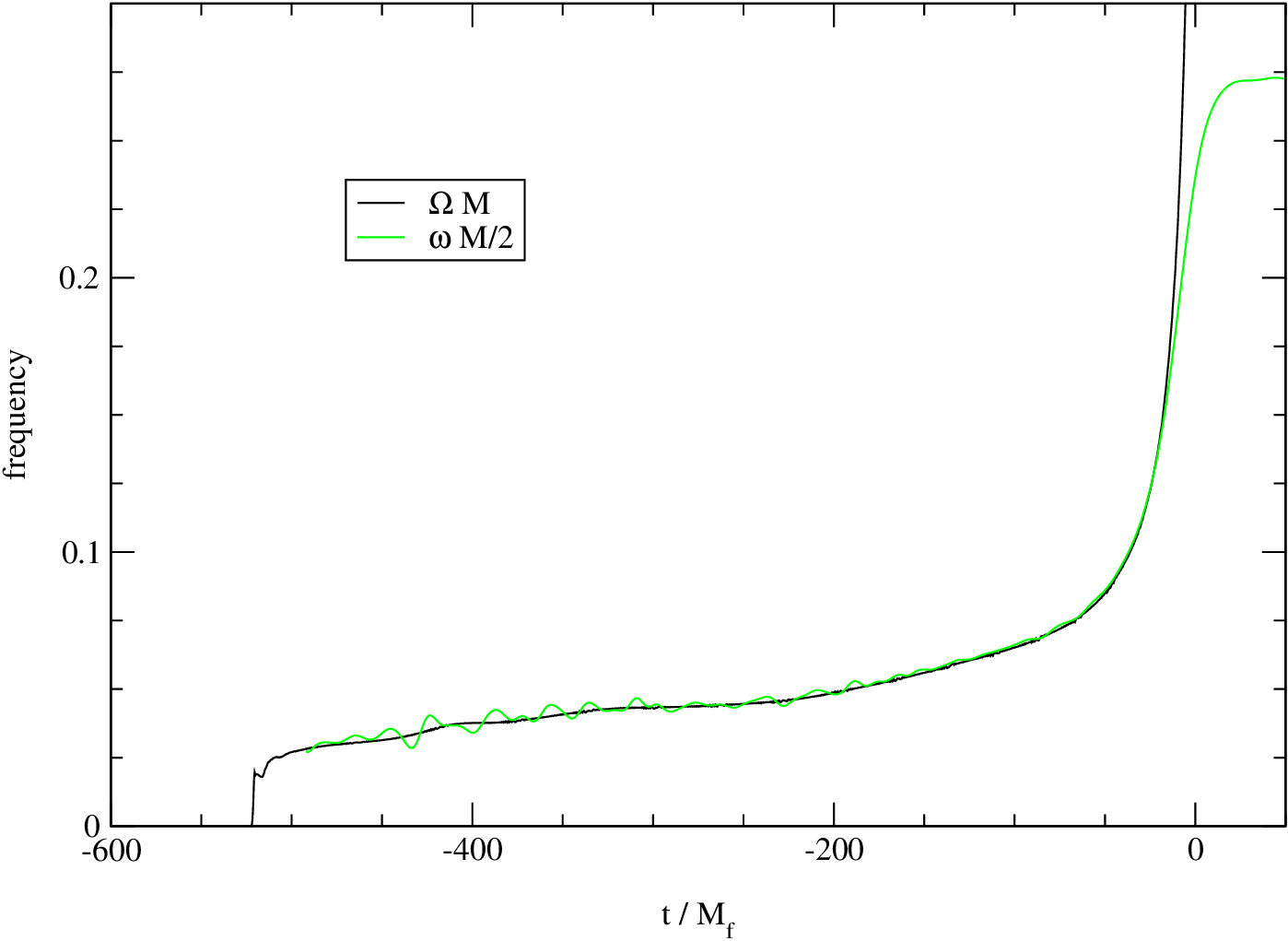}
  \caption{Orbital and radiation frequencies for R4. 
   We show the evolution of the orbital frequency $\Omega$,
   calculated from the coordinate motion of the punctures, compared with the
   wave frequency.  The good correspondence motivates a closer examination 
   of the coordinate-based puncture trajectories. (Note that the radiation
   frequency is divided by two since $\omega = 2\Omega$.)}
  \label{fig:QC9-Freq}
\end{figure}

One way to estimate the utility of such coordinate information is by
comparison with our much more invariant waveform data.  We can, for
instance, compare the coordinate orbital frequency $\Omega$ with the
waveform frequency examined above.  With suitable coordinates, in the
weak-field limit, we would expect the orbital frequency to be
approximately equal to half the gravitational wave frequency.  In
Fig.~\ref{fig:QC9-Freq} we compare the coordinate orbital frequency $
\Omega $ with half the gravitational wave frequency from our R4 run.
We find that, if we shift the orbital freqency data by about $33 M_0$,
the two curves match very well.  Despite some noise in our wave
frequency, most features in the orbital frequency are tracked in the
wave frequency as well.  Exceptions occur at the very beginning of the
simulation, where our coordinates necessarily start with the punctures
non-moving (hence $\Omega=0$) before the brief period through which
the coordinate puncture velocities seem to adapt well to the physical
dynamics.  Likewise, at late times, we see that the orbital puncture
frequency continues to grow while the radiation frequency saturates at
the quasinormal ringing frequency.  This seems to correspond to the
expectation that, at late times, the coordinate motion of the
punctures decouples from the process of radiation generation, with the
punctures continuing to fall into the newly formed black hole while
gravitational ring-down radiation is generated in the final black hole's
pertubative potential barrier region somewhat outside the horizon.

The $33 M_0$ time shift required to realize this agreement can be
interpreted as the time required for the gravitational radiation to
propagate out to the point where it is extracted at $r_{\rm ex}=30 M$.  
We can utilize
this time correspondence to roughly associate phenomena occurring in the
strong field region of the simulations with features in the radiation.  We
have used this time-shift to compare the time at which peak radiation is
generated in our simulations with $T_{merge}$ in
Table~\ref{table:mergertimes}. In our simulations, we find that the merger
occurs about $12M$ before the radiation peak is generated.

\begin{figure}[t]
  \includegraphics*[width=18pc,height=14pc]{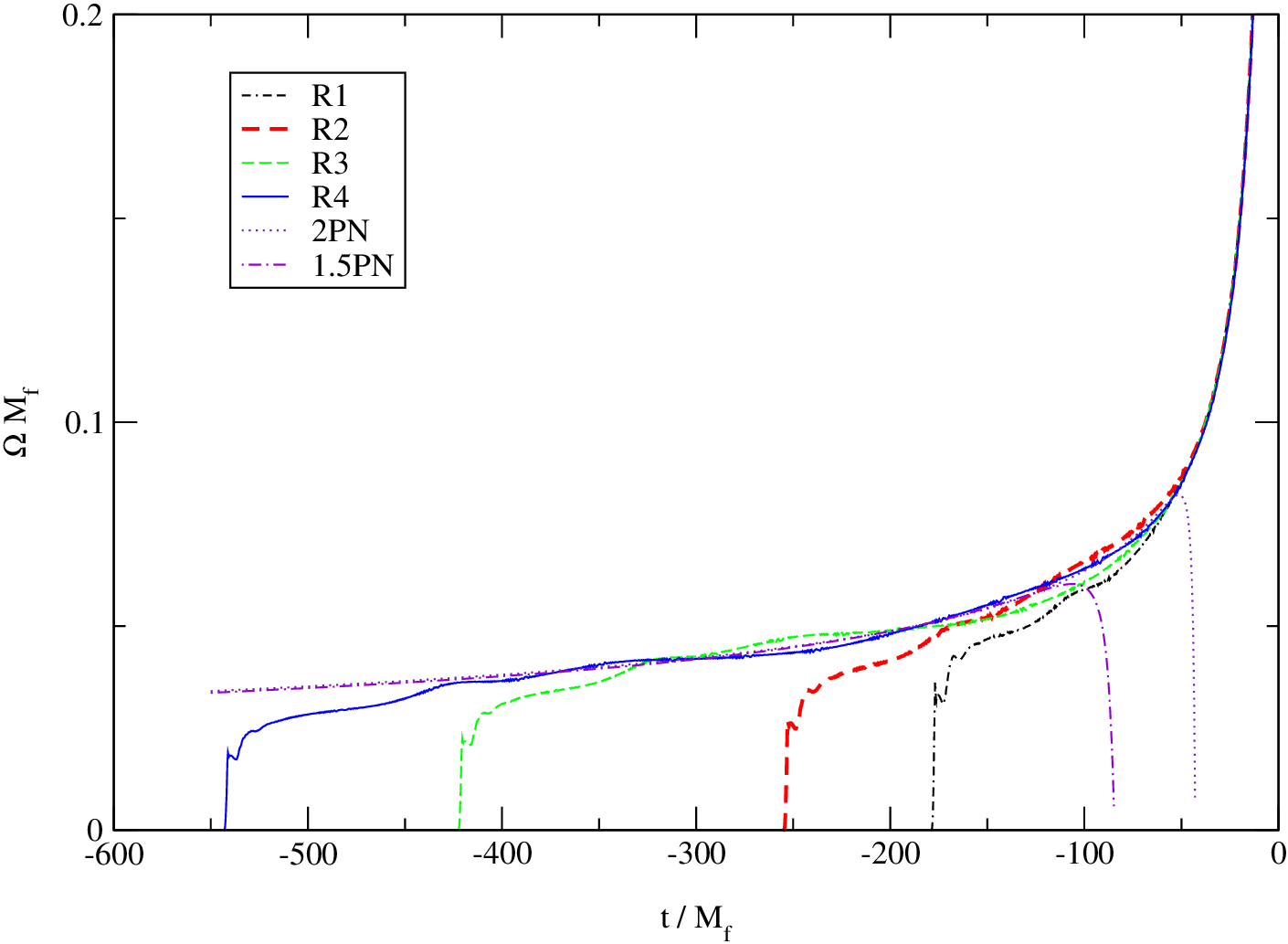}
  \caption{The frequencies as functions of time. 
   Shown are frequencies calculated from
   the waves showing the agreement of the different runs and two
   curves calculated using Post-Newtonian approximations.}
  \label{fig:QCNFreq}
\end{figure}

In Fig.~\ref{fig:QCNFreq} we compare the orbital frequencies of the 
puncture trajectories among our R1-R4 runs.  As was the case with the wave
frequencies we see excellent agreement among the runs after $t=-50 M_f$. Here
we have focused on the earlier region where the results are not quite as
universal.  In each run we notice that after a period of initial angular 
acceleration the orbital frequencies agree to within a few percent.  This 
initial transient period is at least partially a coordinate effect (since
$\Omega$ must begin at $0$), but the strong similarity with the wave-frequencies
suggests that there may be a physical basis for the descrepancies as well.
Particularly for R3 and R4, the shape of the curves suggests a slow
oscillation, perhaps some ellipticity in the motion.  For an external
comparison, we have included the post-Newtonian frequency evolution 1.5PN
and 2PN order, as provided in Ref.~\cite{Blanchet02}, positioned in time so
that all curves agree near $t=-200M_f$. The correspondence with these PN
curves is certainly striking, though we caution that the 2.5PN curve does not
agree so well, with its peak frequency topping out too low for a useful
comparison here.  In subsequent work we intend to explore comparisons with
post-Newtonian calculations in more detail.

\section{Discussion} 
\label{sec:discussion}
We have presented a set of numerical simulations representing
the last few orbits and merger of an equal-mass non-spinning 
binary black hole system.  Though the initial data differ, with the
black holes starting out at separations ranging from $9 M$ to $13 M$, the
calculated waveforms are dominated by universal characteristics.  Over the
period beginning from about $50M$ before the gravitational wave peak, and 
covering approximately the last orbit, all our simulations show profound 
agreement, with differences among the waveforms at no more than about 1\%.
The robustness of this part of our waveform is strong evidence that it
accurately represents the final burst of gravitational radiation from
an astrophysical system of equal-mass, non-spinning black holes, as
predicted by Einstein's field equations.

We also see good agreement among our simulations in the gravitational
radiation generated in the several preceding orbits which we have simulated. 
Excepting an initial transient period about $100M$ for each run, the
earlier portion of the waveforms show good agreement, within approximately 
10\% in phasing and amplitude.  For our longest simulation (R4) this suggests that
we have produced a good approximation of the astrophysical waveform prediction
covering more than $400M$ before the radiation peak.  We are further encouraged
by good agreement over much of this period among the frequency evolution of
our simulated waveforms (especially R4), with the second-order (2PN) 
post-Newtonian predictions.  We are currently exploring the correspondence 
of our waveforms with post-Newtonian calculations, to be developed in more detail 
in a future publication. 

That the late-time part of the simulations (the final orbit and thereafter) 
shows stronger universality than the earlier part of the simulations supports 
idea that the late dynamics of the system is dominated by the strong 
interaction of the holes, and radiative losses, which have the effect of 
reducing dependence on prior conditions.
For the remnant black hole formed in the merger,  
our simulations consistently predict a spin within about $1\%$ of $a/m=0.69.$

Our simulations have employed newly developed numerical relativity
techniques for evolving black holes, which allows the black holes to
propagate accurately across the numerical
domain\cite{Campanelli:2005dd,Baker:2005vv}.  This approach does not
require excision of the black hole interior from the computational
domain.  We have calibrated our approach on one of our black hole
configurations, R1, demonstrating $2^\mathrm{nd}$-order convergence,
and waveform accuracy at the $1\%$ level in the last orbit.  For all
our simulations we have found good energy and angular momentum
conservation as measured by comparing the mass and spin of the final
black hole, measured by its quasinormal ringing, with the expected
remainder after radiative losses.  In future work we expect to
continue to refine our techniques for accuracy and efficiency over
long-lasting simulations.

We have also studied the trajectories traced out by the motion of the black holes
in our numerical coordinate system.  These trajectories suggest a coherent picture of the 
system's evolution which is qualitatively, and in some ways, quantitatively consistent with 
invariant information measured in the radiation waveforms.
This correspondence provides some foundation for giving a tentative
physical interpretation to some coordinate-based information in our
simulations, such as the number and rate of inspiral orbits, and recommends
further research toward a deeper understanding of the properties of the
coordinate systems which we have applied in these simulations.

Our comparative analysis provides some insight into the quality of the initial
data models which we have applied.
We have seen an indication of a small amount (with negligible energy content) 
of spurious radiation arising from initial modeling error.  As generally 
expected the scale of this spurious radiation decreases for increasingly 
well-separated initial data. For sufficiently long-lasting runs this spurious 
radiation is well-segregated in time, limiting its direct significance in 
interpreting the merger radiation.  Of perhaps greater concern are 
indications which suggest ellipticity in the inspiral trajectories and waveforms.
Two aspects of our initial data model may be contributing to this effect.
First, the will be some level of mismatch between the initial separation of the
black holes and the specified initial value of angular momentum.  
We comment, in this regard,  that other simulations we have looked at in preparation for the
work presented here hint that the early transient part of each
simulation, as well as measures such as the merger times, may depend
sensitively on small (1\%) changes in the initial angular momentum.
Secondly, as is customary, the data we have applied have the black holes
set on initially circular trajectories, with vanishing radial momentum components.
This may also be expected to lead to ellipticity, as has been noted in 
binary neutron star simulations~\cite{Miller:2003pd}.
We expect to explore these and similar concerns in future simulations.

Taken together with other recent progress in numerical relativity, these results 
herald a new age of numerical simulations applied to further characterize
and understand strong-field binary black hole interactions and merger radiation.
Future applications will begin to explore the physical parameter space of these 
systems to study, particularly, the effect on the radiation waveforms of 
individual black hole spin, and variations in the black hole mass ratio.

\acknowledgments

We thank David Brown for providing {\tt AMRMG}, and Cole Miller for many 
insightful discussions. This work was supported in part by NASA grants
ATP02-0043-0056 and O5-BEFS-05-0044.  The simulations were carried out using
Project Columbia at NASA Ames Research Center and at the NASA Center for
Computational Sciences at Goddard Space Flight Center. M.K and J.v.M.  were
supported by the Research Associateship Programs Office of the National
Research Council and the NASA Postdoctoral Program at the Oakridge
Associated Universities.

\bibliographystyle{../bibtex/apsrev}

\bibliography{../bibtex/references}

\end{document}